\documentclass{jpsj3}  
\usepackage{txfonts} 
\usepackage[dvipdfmx]{graphicx}

\title{Phase Separation in Kitaev Chain
}
 
\author{Kazuhiro Kuboki \thanks{kuboki@kobe-u.ac.jp}}
\inst{Department of Physics, Kobe University, Kobe 657-8501, Japan
}  
 
\abst{
Kitaev chain is a one-dimensional spinless fermion model that has 
$p$-wave superconducting (SC) states and Majorana zero modes 
at the edge.  Usually this model is analyzed by taking only 
a SC order parameter (OP) into account, but the situation significantly
changes when OPs other than the SCOP are included. 
It turns out that the SC state in the latter case is prone 
to phase separation for  moderate to strong attractiive interactions.

}
 
\begin{document}

\maketitle

\newpage

Kitaev chain is a one-dimensional spinless fermion system which is 
intensively studied as a model to realize $p$-wave superconducting (SC) 
states with Majorana zero modes at the edge. \cite{Kitaev} 
This model is usually analyzed by 
taking only a SC order parameter (OP) into 
account\cite{Minutillo,Cuy,Cuy2,Pozo,Cinnirella}, 
since the original modeling started with a proximity induced 
superconducting state in a nanowire placed on a three dimensional superconductor.

In this study, we consider a one-dimensional spinless fermion model with attractive 
nearest-neighbor (NN)  
interactions by employing a mean-field approximation with Hartree-Fock-Gorkov type decoupling. 
Our purpose is to consider more general situations beyond the modeling of 
Kitaev\cite{Kitaev}, and to show that different behaviors can emerge 
when we include OPs other than the SCOP. 

We treat the model with the Hamiltonian 
\begin{eqnarray}
\displaystyle 
H = & \displaystyle -t\sum_j (a^\dagger_{j+1} a_j + a^\dagger_j a_{j+1})  
-\mu \sum_j a^\dagger_j a_j   \nonumber   \\   
& \displaystyle -V\sum_j a^\dagger_j a_j a^\dagger_{j+1} a_{j+1},
\end{eqnarray}
where $a_j$ is the operator for spinless fermions and $\mu$ denotes 
the chemical potential. 
$t$ and $V$ are the transfer integral and the attractive interaction
between NN sites, respectively. 
We decouple this Hamiltonian by taking the following OPs:  
$p$-wave SCOP 
$\Delta = -i \langle a_{j+1} a_j \rangle$, 
bond OP (BOP) $\chi = \langle a^\dagger_{j+1} a_j \rangle$, and 
fermion density $n = \langle a^\dagger_j a_j \rangle$. 
(Here we denote $n$ as a kind of OP to signal phase separation, 
although it is always finite.) We impose the periodic boundary condition,  
and assume that $\chi$ and $\Delta$ are real.
The self-consistency equations can be obtained by minimizing 
the free energy as follows
\begin{eqnarray}
\displaystyle n = \frac{1}{2} 
- \frac{1}{2N} \sum_k \tanh\Big(\frac{E_k}{2T}\Big)\frac{\xi_k}{E_k}
\end{eqnarray}
\begin{eqnarray}
\displaystyle \chi = 
- \frac{1}{2N} \sum_k \tanh\Big(\frac{E_k}{2T}\Big)\frac{\xi_k}{E_k} \cos k
\end{eqnarray}
\begin{eqnarray}
\displaystyle \Delta = \frac{\Delta}{N}\sum_k  
\tanh\Big(\frac{E_k}{2T}\Big)\frac{V\sin^2 k}{E_k}
\end{eqnarray}
where $\xi_k = -2(t-V\chi) \cos k -\mu -2nV$,  
$\Delta_k = 2V\Delta\sin k$, and $E_k=\sqrt{\xi_k^2+|\Delta_k|^2}$.
$N$ is the total number of lattice sites. 

For numerical calculations we take $t=1$ as the unit of energy, and  
choose $n=0.3$. (Results for other values of $n$ are qualitatively the same.)
In Fig.1, the SCOP ($\Delta$) 
and the BOP ($\chi$) are shown  for several choices of $V$ 
as functions of $T$. 
As seen the superconducting transition temperature $T_C$  naturally  
increases as $V$ becomes large.  

\begin{figure}[htb]
\begin{center}
\includegraphics[width=8.0cm,clip]{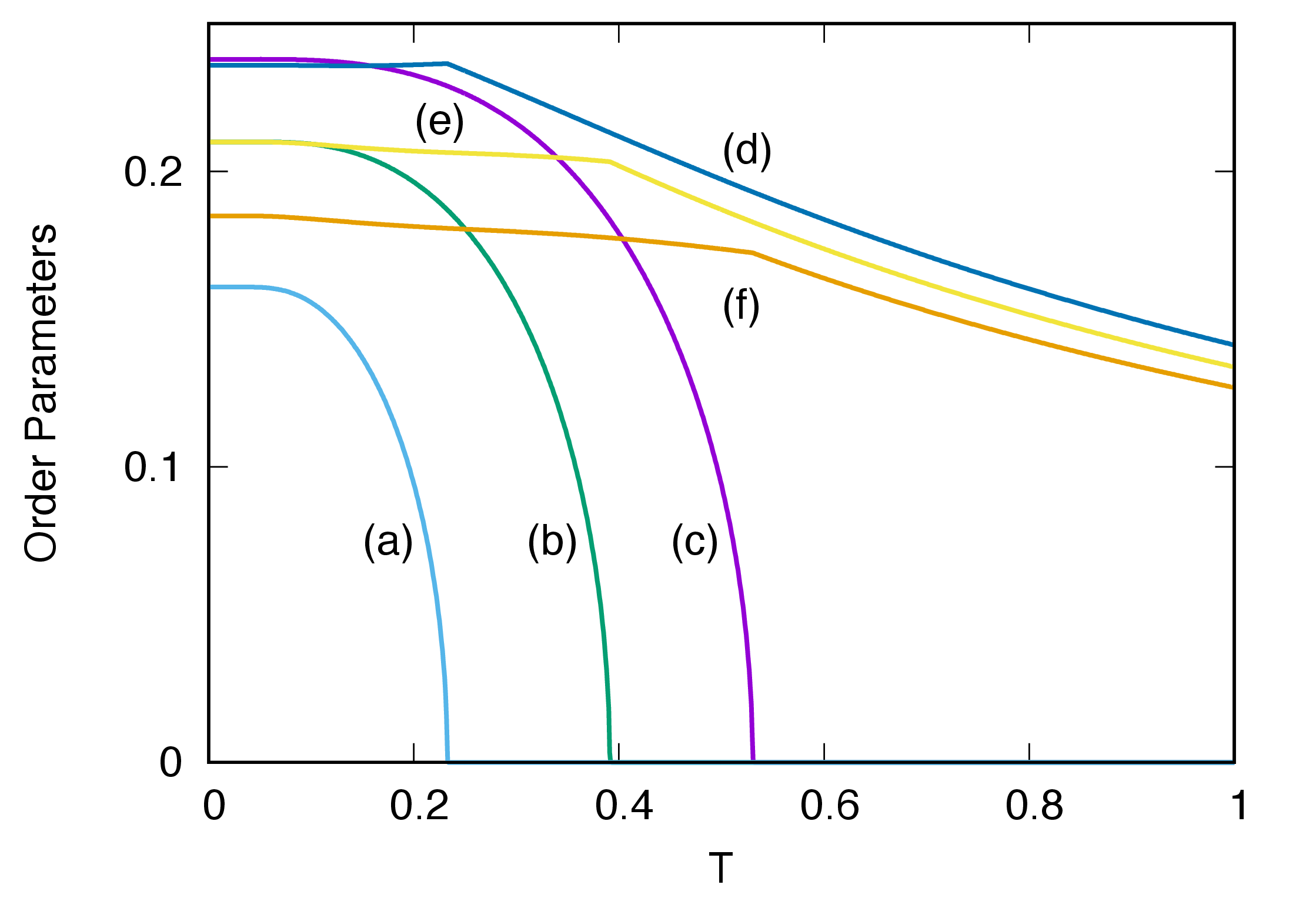}
\caption{(Color online) SCOP $\Delta$ and BOP $\chi$ for 
$t=1$ and $n=0.3$.
(a) $\Delta$ for $V=1.5$, (b) $\Delta$ for $V=2.0$, 
(c) $\Delta$ for $V=2.5$, (d) $\chi$ for $V=1.5$, 
(e) $\chi$ for $V=2.0$, and (f) $\chi$ for $V=2.5$. 
} 
\end{center}
\end{figure}

However, there is a hidden transion to phase separated states 
as we will show by calculating  the compressibility 
$\partial n/\partial \mu$. 
By taking derivatives of Eqs. (3)-(5) with respect to $\mu$, 
the simultaneous equations for $\partial n/\partial \mu$, 
$\partial \chi/\partial \mu$, and $\partial \Delta/\partial \mu$
are obtained as 
\begin{eqnarray}
\displaystyle {\hat A}
\left (\begin{array}{cc} 
\displaystyle 
\frac{\partial n}{\partial \mu} \\
\displaystyle
\frac{\partial \chi}{\partial \mu} \\
\displaystyle
\frac{\partial \Delta}{\partial \mu} 
\end{array}\right )
= 
\left (\begin{array}{cc}
c_1 \\ 
c_2 \\
c_3
\end{array}\right ),   
\end{eqnarray}
where ${\hat A}$ is a $3\times 3$ matrix whose   
matrix elements are given as 
$A_{11} = 1-\frac{V}{N}\sum_k F_1(k)$,
$A_{12} = \frac{V}{N}\sum_k \cos k F_1(k)$, 
$A_{13} = \frac{2V^2\Delta}{N}\sum_k \sin^2k F_2(k) \xi_k$, 
$A_{21} = -A_{12}$,
$A_{22} = 1 + \frac{V}{N}\sum_k \cos^2k F_1(k)$, 
$A_{23} = \frac{2V^2\Delta}{N}\sum_k \cos k \sin^2k  F_2(k) \xi_k$, 
$A_{31} = -\frac{2V^2}{N}\sum_k \sin^2k F_2(k) \xi_k$, 
$A_{32} = \frac{2V^2}{N}\sum_k \cos k \sin^2k F_2(k) \xi_k$, 
and 
$A_{33} = \frac{4V^3\Delta}{N}\sum_k \sin^4k F_2(k)$.
The constants $C_i$ ($i=1,2,3$) are  given as 
$C_1 = \frac{1}{2N}\sum_k F_1(k)$,
$C_2 = \frac{1}{2N}\sum_k \cos k F_1(k)$, and 
$C_3 = \frac{V}{N}\sum_k \sin^2k F_2(k) \xi_k$, 
where the functions $F_1(k)$ and $F_2(k)$ are defined as 
$F_1(k) = \Delta_k^2 \tanh(E_k/2T)/E_k^3 
+\xi_k^2/(2TE_k^2 \cosh^2(E_k/2T))$  
and 
$F_2(k) = 1/(2TE_k^2\cosh^2(E_k/2T)) - \tanh(E_k/2T)/E_k^3$, respectively.
By solving Eq. (5), $\partial n/\partial \mu$, etc. can be obtained.

The $T$ dependences of the compressibility 
are shown in Fig.2.
%
\begin{figure}[htb]
\begin{center}
\includegraphics[width=8.0cm,clip]{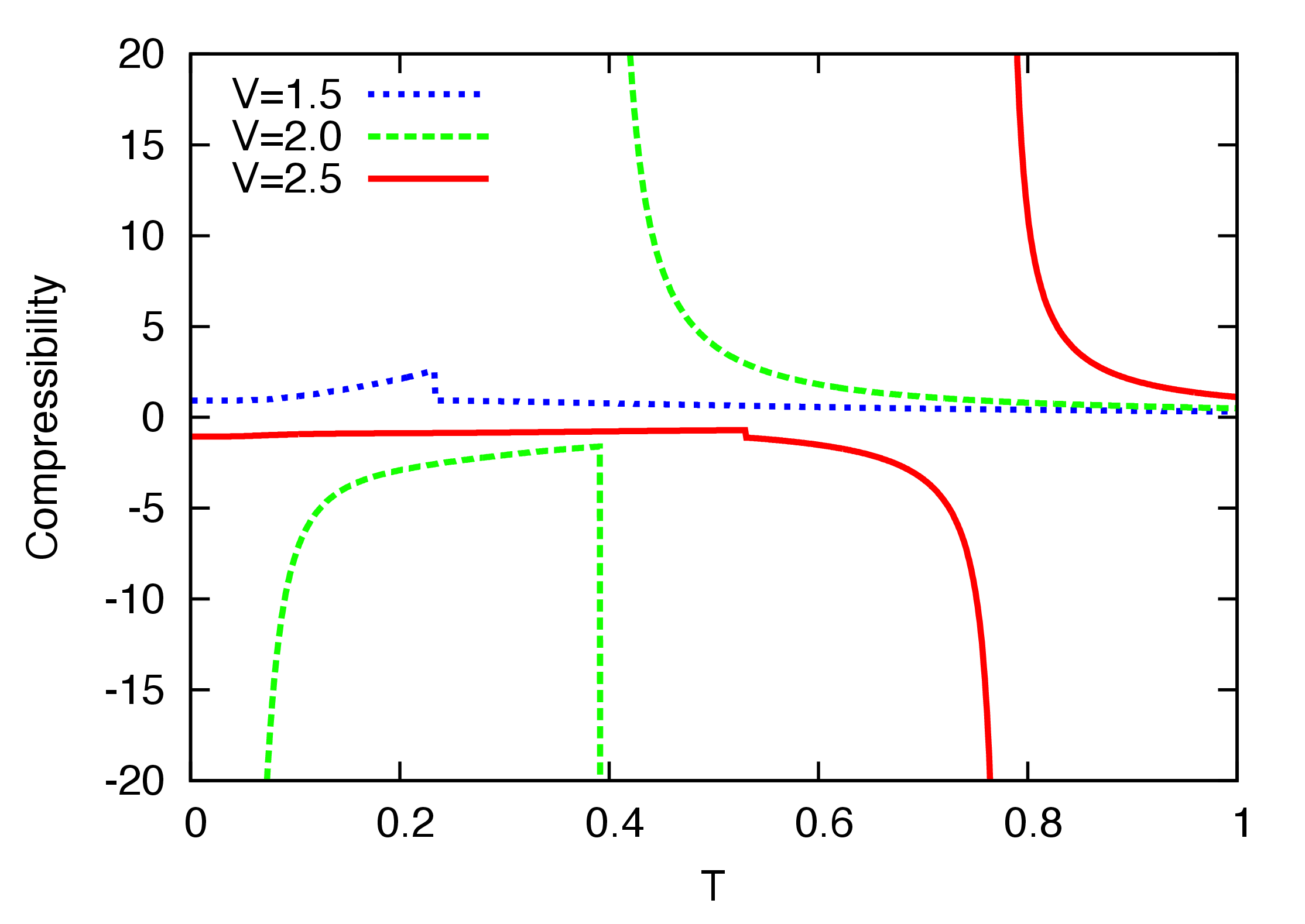}
\caption{(Color online) Temperature dependences of the 
compressibility $\partial n/\partial \mu$ for several choices of $V$ 
with $t=1$ and $n=0.3$.   
} 
\end{center}
\end{figure}
%
For $V=1.5$,  $\partial n/\partial \mu$ is positive 
and finite so that the spatially homogeneous state is stable. 
In the case of $V=2.0$ ($V=2.5$), however, 
$\partial n/\partial \mu$ diverges at $T \sim 0.40$ ($T\sim 0.78$), 
and becomes negative for $T \lesssim 0.40$ ($T \lesssim 0.78$).
These results indicate that there are regions where the homogeneous 
states are not stable and phase separation (PS) occurs. 
In numerical calculations, if we take $n$ as an input parameter 
a spurious solution may be obtained, but the compressibility 
diverges (or becomes negative) in that case. 
When the resultant $\mu$ is taken
as an input parameter, we cannot obtain the original value of
$n$ ($=0.3$), since multiple values of $n$ correspond to a single 
value of $\mu$.  
%
\begin{figure}[htb]
\begin{center}
\includegraphics[width=8.0cm,clip]{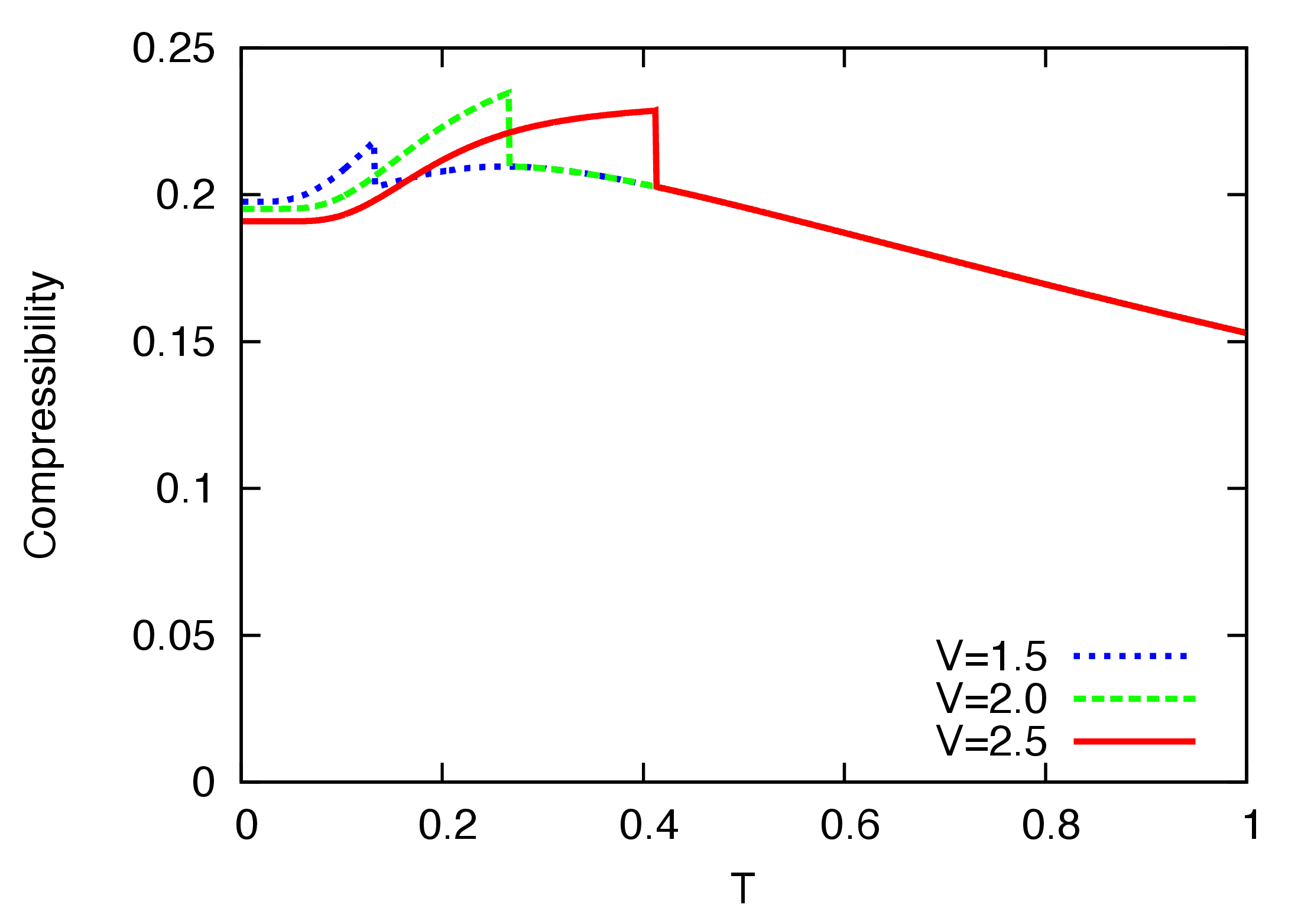}
\caption{(Color online) Temperature dependences of the 
compressibility $(\partial n/\partial \mu)_0$ 
when OPs other than the SCOP are omitted.
Parameters are the same as in Fig.2.
} 
\end{center}
\end{figure}
%

For comparison, we calculate the compressibility when 
only the SCOP is taken into account.
This amounts to replace $\xi_K$ with $-2t\cos k -\mu$,
and we denote the compressibility in this case as 
 $(\partial n/\partial \mu)_0$.
In Fig.3, $T$ dependences of $(\partial n/\partial \mu)_0$ are shown. 
It is seen that $(\partial n/\partial \mu)_0$ is always positive 
and finite so that PS is absent.
The temperature at which $(\partial n/\partial \mu)_0$ has a peak 
corresponds to $T_C$ for each $V$.
Here we note that $T_C$ in this case is different from those in Fig.1, 
because the dispersion $\xi_k$ is diffrent.  

The above considerations indicate that the occurrence of 
PS is due to the OPs usually ignored. 
If we discard the term $2nV$ and retain the term $V\chi$ in $\xi_k$, 
PS would not occur.
This means that the former term, which is  proportional to the particle 
density in the dispersion relation $\xi_k$, is the primary cause of PS. 
On the contrary, if we drop the term $V\chi$ and keep the term $2nV$, 
PS may occur but the value of $V$ necessary to induce PS becomes larger. 
This is because the inclusion of $V\chi$ in $\xi_k$ reduces the band width, 
and so the density of states at the Fermi level is increased, leading to 
the situation favorable to the phase transition. 
Then dropping the term $V\chi$ works to suppress PS. 

Similar behavior is also observed in strongly correlated 
electron systems; 
the two-dimensional $t-t'-J$ model treated within a slave-boson mean-field 
approximation\cite{OF,LNW} is an example. 
In this case, the dispersion relation of spinons depends on the 
hole density and the BOP, and the contributions from their derivatives with
respect to the chemical potential may lead to the divergence of 
the compressibility,  and hence phase separation.\cite{tJkk}

In real systems, there exist long-range (or screened) Coulomb interactions 
(LRC) between electrons, and they work to suppress phase separation. 
It might be expected that the competition between the tendency toward 
inhomogeneous states and the LRC would result in charge density wave states 
with finite wave vectors ${\bf q}$, instead of ${\bf q}={\bf 0}$ in the case of PS, 
as a compromise between the two. 
This will be an intriguing future problem.

In summary, we have examined the stability of the SC state in Kitaev chain 
and found that it is prone to phase separarion when the attractive 
interaction becomes strong. 
Phase separation in this case is caused by the particle-density dependences 
of the dispersion relation, which are arising from the OPs other than the SCOP 
but are often ignored.  
This issue may also be important for the stability of the superconducting 
states in other kinds of models.

\medskip
\begin{acknowledgment}
 The author thanks M. Hayashi for useful discussions. 
 This work was supported by JSPS KAKENHI Grant Number 23K04525. 
 
\end{acknowledgment}


\end{document}